\documentclass[twocolumn,aps,prl,superscriptaddress]{revtex4}
\usepackage{graphicx}
\usepackage{bm}

\def\bea{\begin{eqnarray}}
\def\eea{\end{eqnarray}}
\def\be{\begin{equation}}
\def\ee{\end{equation}}

\begin{document}

\title{Precision Test of Mass Ratio Variations with Lattice-Confined Ultracold Molecules}

\author{T. Zelevinsky}
\affiliation{JILA, National Institute of Standards and Technology
and University of Colorado, Boulder, CO 80309-0440, USA}
\author{S. Kotochigova}
\affiliation{Physics Department, Temple University, Philadelphia, PA
19122-6082, USA}
\author{J. Ye}
\affiliation{JILA, National Institute of Standards and Technology
and University of Colorado, Boulder, CO 80309-0440, USA}

\begin{abstract}     
We propose a precision measurement of time variations of the
proton-electron mass ratio using ultracold molecules in an optical
lattice. Vibrational energy intervals are sensitive to changes of
the mass ratio.  In contrast to measurements that use
hyperfine-interval-based atomic clocks, the scheme discussed here is
model-independent and does not require separation of time variations
of different physical constants.  The possibility of applying the
zero-differential-Stark-shift optical lattice technique is explored
to measure vibrational transitions at high accuracy.

\end{abstract}


\date{\today}
\maketitle

\newcommand{\w}{3.25in}

\newcommand{\MorseFigure}[1][\w]{
\begin{figure}
\includegraphics*[width=2.6in]{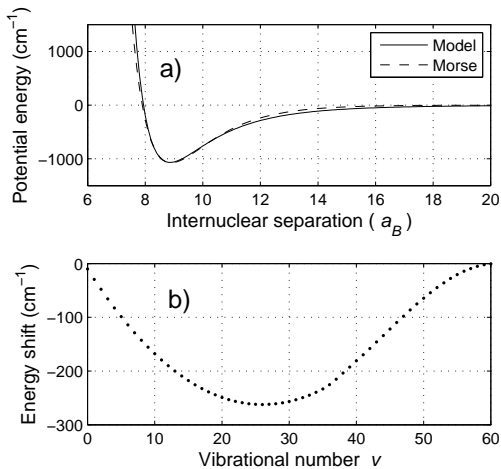}\hfill
\caption{(a) The model potential curve for the Sr$_2$ ground
state (solid line), and the Morse potential fitted to three
parameters (dashed line). (b) Vibrational energy sensitivities
to $\Delta\mu/\mu$,
as a function of the vibrational number $v$.
} \label{fig:Morse}
\end{figure}
}

\newcommand{\RamanSchemeFigure}[1][\w]{
\begin{figure}
\includegraphics*[width=2.17in,height=2.3in]{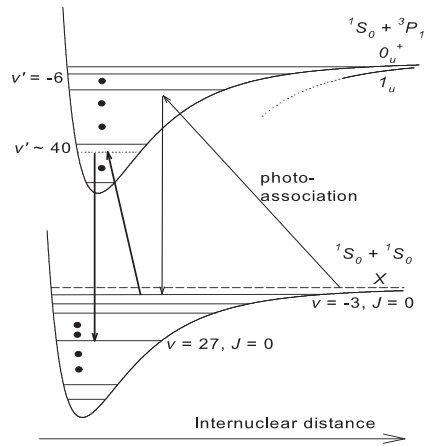}\hfill
\caption{The proposed Scheme I for precision Raman spectroscopy of
Sr$_2$ ground state vibrational level spacings. A two-color
photoassociation pulse prepares molecules in the $v=-3$ vibrational
level.  Subsequently, a Raman pulse couples $v=-3$ and $v=27$, via
$v'\sim 40$ of $0_u^+$.} \label{fig:RamanScheme}
\end{figure}
}

\newcommand{\FCFigure}[1][\w]{
\begin{figure}
\includegraphics*[width=2.6in]{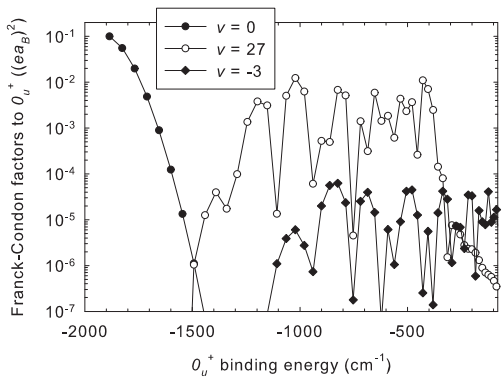}\hfill
\caption{Calculated Franck-Condon factors, or transition dipole moments
squared, between vibrational levels of the $X$ and $0_u^+$ electronic
states as a function of the binding energy for vibrational levels of
$0_u^+$ ($J=0$ and $J'=1$).  As relevant to the text, only the FCFs
of $v=0$, 27, and $-3$ are shown.} \label{fig:FC}
\end{figure}
}

\newcommand{\MagicFreqsFigure}[1][\w]{
\begin{figure}
\includegraphics*[width=2.6in]{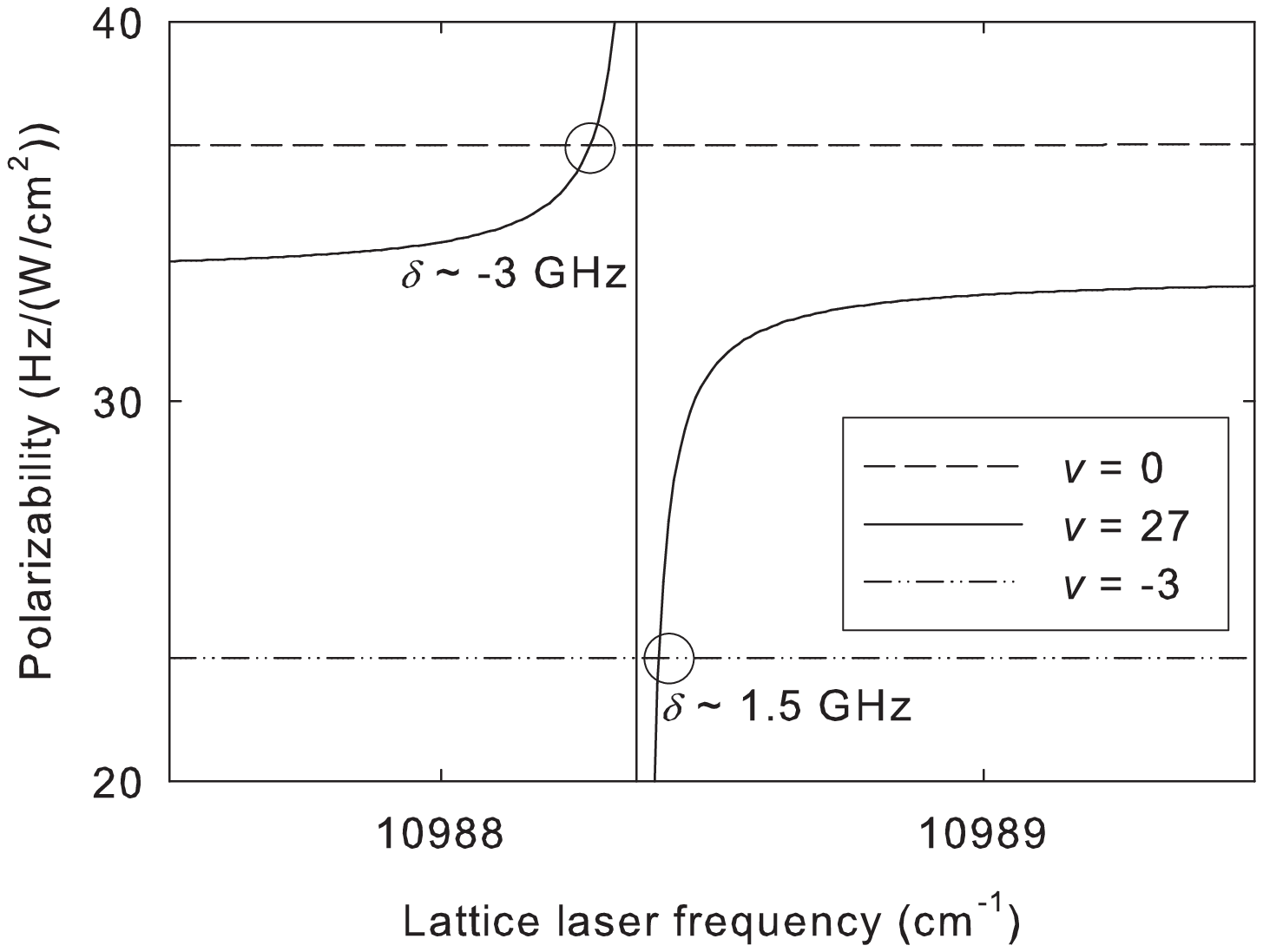}\hfill
\caption{Magic frequencies for the optical lattice near 910 nm
(10990 cm$^{-1}$). The detunings from resonance are
$\sim$ 1.5 and $-3$ GHz for Schemes I and II, respectively, and
must be confirmed experimentally.}
\label{fig:MagicFreqs}
\end{figure}
}

Ultracold molecules open new opportunities for precision
measurements of possible variations of fundamental physical
constants.  The test of time variation of the proton-electron mass
ratio $\Delta\mu/\mu$ ($\mu\equiv m_p/m_e$, where $m_p$ and $m_e$
are the proton and electron masses) is particularly suitable, since
molecules are bound by electronic interactions while ro-vibrations
are dominated by nuclear dynamics. Recent proposals to search for
$\Delta\mu/\mu$ include detecting changes in the atomic scattering
length near a Feshbach resonance \cite{FlambaumChinMassRatio} or
using near-degeneracies of molecular vibrational levels from two
different electronic potentials to probe microwave frequency shifts
arising from $\Delta\mu/\mu$ \cite{DeMilleThesis,FlambaumDiatomic}.
We propose a two-color (Raman) optical approach to directly
determine vibrational energy spacings within a single electronic
potential of ultracold dimers in an engineered optical lattice. The
measurement relies on the cumulative effect of $\Delta\mu/\mu$ on
the excited vibrational levels, and utilizes the entire molecular
potential depth to enhance precision by choosing two vibrational
levels with maximally different sensitivity.

For a given physical system, we can write a proportionality relation
between $\Delta\mu/\mu$ and the corresponding fractional change in
transition frequency $\Delta\nu/\nu$ as
\be
\frac{\Delta\mu}{\mu}=\kappa\frac{\Delta\nu}{\nu}.
\label{eq:GeneralEnhanFact}
\ee
Only dimensionless quantities appear
in Eq.~(\ref{eq:GeneralEnhanFact}), to avoid introducing a time
dependence of the Cs hyperfine frequency, for example.  In the case
discussed here, $\kappa$ is of order one and has
a small potential-dependent uncertainty.  The fractional uncertainty $\delta\mu/\mu$
of the measurement of $(\Delta\mu\pm\delta\mu)/\mu$ must be
minimized. For a given frequency measurement uncertainty
$\delta\nu$, from Eq.~(\ref{eq:GeneralEnhanFact}) we obtain \be
\frac{\delta\mu}{\mu}=\kappa\frac{\delta\nu}{\nu}=
\frac{d\ln\mu}{d\ln\nu}\frac{\delta\nu}{\nu}=
\left(\frac{d\nu}{d\ln\mu}\right)^{-1}{\delta\nu}.
\label{eq:FracMuError} \ee The last step in
Eq.~(\ref{eq:FracMuError}) is motivated by the assumption that
experimental limitations constrain $\delta\nu$ rather than
$\delta\nu/\nu$.  Eq.~(\ref{eq:FracMuError}) indicates that the
quantity $(d\nu/d\ln\mu)$ must be maximized. In other words, we
search for the energy gap with the maximum absolute frequency shift
arising from a given fractional mass ratio change. While a microwave
measurement \cite{DeMilleThesis,FlambaumDiatomic} could have a
smaller $\delta\nu$ than the optical frequency-comb-based
Raman approach, the latter
maximizes sensitivity through the cumulative effect of the entire
molecular potential depth.

Heteronuclear dimers may be advantageous for the present proposal
due to deeper electronic ground state potentials by a factor of
$\sim$2-5 (${d\nu}/{d\ln\mu}$ is proportional to the same factor).
On the other hand, homonuclear dimers can lead to higher precision
since their radiatively long-lived vibrational levels in the
electronic ground state are insensitive to blackbody radiation.
Molecules based on even isotopes of
alkaline-earth-type atoms (e.g. Sr, Ca, Yb) enjoy the lack of hyperfine and
magnetic structure in the electronic ground state,
simplifying the preparation of the system and reducing systematic shifts.

As in the work on a Sr-atom optical lattice clock
\cite{BoydClock07} and narrow-line
photoassociation (PA) spectroscopy \cite{tz}, we propose to use
ultracold Sr dimers confined in an optical lattice for the
$\Delta\mu/\mu$ experiment. Vibrationally excited Sr$_2$ in the
electronic ground state is produced via PA. Raman spectroscopy aided
by a femtosecond optical frequency comb will be used to interrogate the
energy spacings between deeply bound vibrational levels and those
closer to the dissociation limit. The Franck-Condon factors (FCFs)
between the electronic ground state $X$ potential (dissociating to
$^1S_0+^1S_0$) and the excited $0_u^+$ potential (dissociating to
$^1S_0+^3P_1$, with $ungerade$ symmetry and the atomic angular
momentum projection onto the molecular axis $\Omega=0$) are
sufficiently favorable to enable Raman transitions between two
vibration levels in the $X$ state of Sr$_2$ via $^3P_1$, as both
potentials are dominated by van der Waals interactions. Since the
$X$ potential is 30 THz deep
\cite{Sr2DepthExpt80,Sr2DepthExpt84,ShortRangeSrCPL03}, and the
relative stability of the Raman lasers can be maintained to better
than 0.1 Hz via the comb \cite{LudlowNarrowLinewidth}, we expect a
precision of better than $\sim$ (0.1 Hz)/(10 THz) $=10^{-14}$ in the
test of $\Delta\mu/\mu$.

Other tests based on atomic frequency metrology constrain
$\Delta\mu/\mu$ to $\sim 6\times 10^{-15}$/year
\cite{KarshClocksVariation}, and the recent evaluation of
astronomical NH$_3$ spectra limits $\Delta\mu/\mu$ to $\sim 3\times
10^{-16}$/year \cite{FlambaumNH3}. The atom-based tests rely on
theoretical interpretations such as the Schmidt model, since
electronic and fine structure transitions do not directly depend on
$\mu$, and hyperfine transitions simultaneously depend on $\mu$,
$\alpha$ (the fine structure constant), and $R_{\infty}$ (the
Rydberg constant).  The NH$_3$ result is based on molecular lines
and is therefore less model-dependent, but relies solely on
cosmological observations. In addition, it disagrees with the
H$_2$-based result relevant to the same cosmological age ($10^{10}$
years) that indicates non-zero $\Delta\mu/\mu$ at the
$10^{-15}$/year level \cite{UbachsPRL06}. The molecular system
proposed here provides a direct test of present-day variations
with a competitive precision
and a weak dependence on theoretical modeling \cite{Chardonnet}.

\MorseFigure
In this work we select vibrational levels of the ground state
potential that have the largest and smallest sensitivities to $\Delta\mu/\mu$.
To model the Sr$_2$ ground state potential $V(r)$,
we combine the experimental RKR potential
\cite{Sr2DepthExpt84} with its long-range dispersion form $-c_6/r^6-c_8/r^8$, where
$c_6 = 3100$ $E_ha_B^6$ \cite{PorsevDerevianko06}, $c_8 = 1.9\times 10^5$
$E_ha_B^8$ is determinted by smoothly connecting to the RKR potential,
and $r$ is the internuclear separation in units of $a_B$
($E_h=4.36\times 10^{-18}$ J and $a_B=0.0529$ nm).
The model potential has 61 vibrational levels, depth $d=4.7\times 10^{-3}$ $E_h$,
the minimum at the internuclear separation $r_0 = 8.9~a_B$, and a scattering length
of $\sim 8~a_B$ \cite{MickelsonPRL05,YasudaPRA06}.
The Morse potential can be quantized analytically and is a convenient approximation
to the ground state molecular potential,
\be
V_M(r)=d(1-e^{-a(r-r_0)})^2-d,
\label{eq:Morse}
\ee
where $a\approx 0.7$ $a_B^{-1}$ for Sr$_2$.
The Morse energy levels are
\be \epsilon_n =
2\epsilon_0(n+\frac{1}{2})-\epsilon_0^2(n+\frac{1}{2})^2/d-d,
\label{eq:MorseSpectrum}
\ee
where $\epsilon_0$ is approximately the
zero-point energy.
Note that the Morse spectrum is valid only if $\epsilon_n-\epsilon_{n-1}>0$, or
$n<d/\epsilon_0$, which means that $V_M(r)$ has about $N\sim 40$ bound levels.
Figure \ref{fig:Morse} (a) compares the Sr$_2$ model potential and its Morse approximation.

Since $\epsilon_0\propto \mu^{-1/2}$, the logarithmic derivative of
the expression for the $n$th vibrational energy level is
\be
\frac{d\epsilon_n}{d\ln\mu}=-\epsilon_0(n+\frac{1}{2})+\epsilon_0^2(n+\frac{1}{2})^2/d.
\label{eq:LogDeriv}
\ee
Equation (\ref{eq:LogDeriv}) is the energy
level sensitivity to a fractional mass ratio change, with the
maximum absolute sensitivity for $n_{\rm{max}}\simeq N/2$, and lowest
sensitivity near the bottom and the top of the potential well, as expected
given a fixed potential depth.  The sensitivities were also determined for our
Sr$_2$ model by calculating vibrational energies for slightly different atomic masses.
Both the Morse approximation and more realistic calculation
point to $25\lesssim v\lesssim 28$ as the most sensitive.
Figure \ref{fig:Morse} (b) shows the level sensitivities to $\Delta\mu/\mu$.
We choose to work with $v=27$, and the $\Delta\mu/\mu$ measurement is optimized if
either a weakly bound or the deepest vibrational level is chosen as
the reference level for Raman spectroscopy.

\RamanSchemeFigure
Based on considerations of
sensitivity as well as molecular transition strengths, we propose
two schemes for the measurement of $\Delta\mu/\mu$.  While they
yield the same information, their combination provides a
consistency check on the experimental method.  The first scheme is
more straightforward as it relies on one of the
least-bound vibrational levels for the Raman transition.  The second
scheme involves an additional Raman step to drive the weakly bound
molecules into a deeper vibrational level. Below, the rotational
angular momenta are $J=0$ for $X$ and $J'=1$ for $0_u^+$ and $1_u$.
The transition strengths are obtained from relativistic configuration
interaction $ab$ $initio$ calculations \cite{KotochJCP,KotochTiesingaPRA06}
adjusted to agree with measurements
of weakly bound $0_u^+$ vibrational levels \cite{tz}.

\FCFigure
Scheme I is illustrated in Fig. \ref{fig:RamanScheme}, and
measures the energy difference between the weakly bound $v=-3$
(negative vibrational numbers imply
counting from the top of the potential with the least-bound level
being $-1$) and $v=27$, the latter being most sensitive to mass
ratio changes.  Step Ia is two-color PA into $v=-3$ via $v'=-6$
\cite{tz} using 689 nm light, with the two colors detuned by about 10
GHz (the primes refer to vibrational levels of the $0_u$ excited
electronic potential). Step Ib is three-level Raman spectroscopy,
$v=-3\rightarrow v'\sim 40\rightarrow v=27$.
Figure \ref{fig:FC} shows the FCFs, that include the transition dipole
moments and are defined as $|\langle v,J|e\vec{{\bf r}}|v',J'\rangle|^2$,
for $v=0$, $v=27$, and $v=-3$ to any vibrational level of $0_u^+$.
For $v=-3$ the FCFs
approach $10^{-4}$ $(ea_B)^2$ for a number of $0_u^+$ vibrational
levels with binding energies smaller than 1000 cm$^{-1}$, where $e$
is the electron charge. For $v=27$ the maximum FCFs are about
hundredfold larger, and quickly decrease for binding energies
smaller than 400 cm$^{-1}$. This suggests using $0_u^+$ intermediate
levels with binding energies around 400 cm$^{-1}$ to balance the
Raman transition strengths, such that the laser wavelengths are
in the 700-750 nm range.
The resulting FCFs of $\sim 10^{-4}$ $(ea_B)^2$ imply that
for a 1 GHz detuning from the intermediate level $v'\sim 40$ and
laser intensities of 2 W/cm$^2$, the two-photon Rabi rate is $\sim
2\pi\times 10$ Hz.

Scheme II that probes deeper levels is possibly less sensitive to
collisional relaxation.  Its disadvantage is an extra step of
two-photon population transfer. Step IIa is the same as Ia.
Step IIb is Raman population transfer
$v=-3\rightarrow v'\sim 40\rightarrow v=27$, analogous to Step Ib.
Finally, step IIc is $v=27\rightarrow v'\sim 10\rightarrow v=0$
Raman spectroscopy, with the wavelengths in the 700-800 nm range. As
shown in Fig. \ref{fig:FC}, the FCFs between $v=0$ and the
vibrational levels of $0_u^+$ are very large near the bottom of $0_u^+$
and decrease rapidly for levels with binding energies smaller than
1500 cm$^{-1}$. This suggests the choice of $v'\sim 10$ for the
intermediate level as it balances the FCFs within the Raman
transition to $\sim 10^{-6}$ $(ea_B)^2$. For these transition
strengths, and using 50 MHz detunings from $v'\sim 10$ and Raman
laser intensities of 10 W/cm$^2$, the Rabi rate for step IIc is
$\sim 2\pi\times 10$ Hz.

The experiment critically depends on the control over
systematic effects and the ability to perform Raman
spectroscopy on a large number of molecules for a good signal to
noise ratio. Using an optical lattice to trap the
ultracold molecules \cite{tz} is beneficial both for attaining high
densities on the order of $10^{12}$/cm$^3$, and for controlling
systematic shifts.  The zero-differential-Stark-shift (or $magic$
$frequency$) technique allows precise and accurate neutral-atom
clocks \cite{KatoriProposal03,LemondeClock06,BoydClock07}. It relies
on the crossing of dynamic polarizabilities of the two probed states
at a certain lattice frequency. Such a lattice ensures a vanishing
differential Stark shift and a suppression of
inhomogenous Stark broadening.  For PA at the $^1S_0+^3P_1$ dissociation limit,
the polarization-dependent magic wavelength is near $914$ nm (10950 cm$^{-1}$).

\MagicFreqsFigure
Analogously, we can search for a
zero-differential-Stark-shift lattice frequency for the proposed
pairs of Sr$_2$ vibrational levels of the $X$ potential.
The Stark shifts of vibrational levels are proportional to dynamic
polarizabilities \cite{KotochJCP,KotochTiesingaPRA06}, and are
independent of the light polarization for $J=0$.  The polarizability
of Sr$_2$ in the vicinity of 914 nm slowly decreases with vibrational
quantum number of $X$.
However, tuning the lattice frequency to near
resonance with a vibrational transition from
$X$ to 1$_u$ makes it possible to match
the polarizabilities of two vibrational levels in $X$. As shown in
Fig.~\ref{fig:MagicFreqs}, our calculations indeed reveal lattice frequency values
in the vicinity of 11000 cm$^{-1}$ for which the polarizabilities
of $v=27$ and $v=-3$, as well
as those of $v=0$ and $v=27$, are equal.  The precise location of the resonance
will shift as more experimental input becomes available.

A possible disadvantage of working near these resonances is enhanced
scattering of lattice light. We estimate the scattering rate from
the resonance strength as well as from the lattice detuning needed
to achieve the zero-differential-Stark-shift condition.  From our
model, these detunings are $\sim 1.5$ GHz for Scheme I
and $\sim -3$ GHz for Scheme II.  Calculations show that the total
spontaneous decay rate of $1_u$ levels with $4000$ cm$^{-1}$ binding
energy (as in Fig.~\ref{fig:MagicFreqs}) is $2\pi\gamma$ = $2\pi\times 9$ kHz. The effective photon
scattering rate is given by $\Gamma_s\equiv 2\pi\gamma_s=(\pi
s\gamma)/(2\delta/\gamma)^2$, where $\delta$ is the detuning from
the resonance. The measure of the lattice intensity $I$ is $s$, such
that $s=I/I_{\rm{sat}}$, with the effective saturation intensity for
a single vibrational channel
$I_{\rm{sat}}=(\pi c\hbar^2\gamma^2)/(4f_{\rm{FC}})$, where
$2\pi\hbar$ is the Planck constant and $f_{\rm{FC}}$ is the FCF
($f_{\rm{FC}}\sim 5\times 10^{-5}$ $(ea_B)^2$ for the resonance in
Fig.~\ref{fig:MagicFreqs}). The estimated scattering rates are thus
$\sim 4/$s for Scheme I and $\sim 1/$s for Scheme II, for $I$ = 10
kW/cm$^2$.  This intensity supports a trap a few $\mu$K deep. Note
that the scattering of lattice photons due to $0_u^+$ vibrational
levels, as well as non-resonant $1_u$ levels, is
strongly suppressed.

Further, we estimate the incoherent scattering rate of Raman
lasers by the intermediate $0_u^+$ levels.  For the $v=-3\rightarrow v'\sim 40$
transition, for example, this scattering rate is $<0.1$/s for a 1 GHz detuning and
2 W/cm$^2$ laser intensity.  For the $v=27\rightarrow v'\sim 10$ transition,
it is $\sim 1$/s for a 50 MHz detuning and 10 W/cm$^2$ intensity.
These estimates are conservative since the excited state population is expected to be
suppressed when the Raman condition is fulfilled.  Thus the scattering rates of the
lattice and spectroscopy lasers will not limit the $\sim 10$ Hz power-broadened
linewidth of the two-photon transition.

The Stark shift of $v$ due to one of the Raman lasers is
$\Delta=(s\gamma^2)/(8\delta)=\gamma_s(\delta/\gamma)$
in the large detuning limit.
We estimate the near-resonant
contributions to Stark shifts to be $\sim 50$ Hz for both schemes
for each Raman laser.  Moreover, the shifts of the two vibrational
levels within a Raman pair have the same sign which leads to
cancellation with the proper power balance of the Raman beams. In
addition to the near-resonant shift, there is a background shift due
to other molecular levels (which can be partially compensated
by slightly shifting the power balance in the Raman beams).
In the 700-800 nm wavelength range and
for Raman laser intensities of $\sim 10$ W/cm$^2$, the background
Stark shifts are $\sim 100$ Hz. Hence, in the worst case the Raman
beam intensity must be controlled to $<1\%$ for a sub-Hz measurement.

Other major systematic effects are magnetic field fluctuations,
and Sr density variations in the lattice.  Our past work on the
$^{88}$Sr atomic clock \cite{Sr88Clock05} demonstrates that these
effects can be controlled to below 1 Hz. Importantly, the absence of
magnetic structure in the ground electronic state of Sr$_2$ with
$J=0$ should significantly reduce any magnetic shifts.

To experimentally search for the $X$
vibrational levels, we will proceed in two steps. The first is to
locate the $v=-3$ level of $X$ as follows.
PA into $v'=-6$ leads to preferential decay into $v=-3$ \cite{tz}.
Tuning a laser to $v=-3\rightarrow v'=-6$ will
convert the $v=-3$ molecules back to atom pairs. Hence, observation of a
reduced atom trap loss associated with PA should allow locating the
$v=-3$ level.  Once the $v=-3$ level is identified, similar methods
can be used to find the $v'\sim 40$ level of $0_u^+$ and the $v=27$
level of $X$, as well as the other proposed levels, in a
stepwise manner. For example, if the molecules are transferred from
$v=-3$ to $v'\sim 40$, a reduced fraction of Sr$_2$ will be
converted back to atoms in the procedure outlined above. The
lifetime of the $v=-3$ molecules should be long due to the lack of
radiative decay channels.
The collisional relaxation rates are not well known and must be
determined experimentally.

In conclusion, our analysis of Sr$_2$ dimers shows that ultracold
non-polar molecules in a zero-differential-Stark-shift optical
lattice is an excellent system for measuring time variations of mass
ratios. This system provides a model-independent test that is based
on different physics than atomic clocks. We expect a sub-Hz
frequency measurement for a $\sim 10^{-14}$/year test of
$\Delta\mu/\mu$, with future improvements in precision by at least a
factor of ten.

We thank D. DeMille, P. Julienne, A. Derevianko, M. Boyd, and A. Ludlow for valuable
discussions. We acknowledge NSF, NIST, DOE, and ARO for support.


\end{document}